\documentclass[12pt,a4paper]{article}
\usepackage[latin1]{inputenc}
\usepackage[left=20.00mm, right=20.00mm, top=20.00mm, bottom=20.00mm]{geometry}
\usepackage{graphicx,multirow,listings,supertabular,booktabs,subcaption,boldline,enumitem,dblfloatfix}
\usepackage[]{algorithm2e}
\usepackage[noadjust]{cite}
\usepackage[svgnames]{xcolor}
\usepackage{eso-pic} 
\definecolor{diffstart}{named}{Grey}
\definecolor{diffincl}{named}{Green}
\definecolor{diffrem}{named}{OrangeRed}
\definecolor{light-gray}{gray}{0.80}

\lstdefinelanguage{diff}{
	basicstyle=\ttfamily\normalsize,
	morecomment=[f][\color{diffstart}]{@@},
	morecomment=[f][\color{diffincl}]{+\ },
	morecomment=[f][\color{diffrem}]{-\ },
}

\usepackage{tikz}

\usepackage{balance}

\lstdefinestyle{mystyle}
{
	frame=single,
	language=bash,
	basicstyle=\footnotesize\ttfamily,
	keywordstyle=\bfseries,
	showstringspaces=false,
	otherkeywords={11,ff,dd,aa}
}
\makeatletter
\def\lst@makecaption{%
	\def\@captype{table}%
	\@makecaption
}
\makeatother

\author{Muhammad Abdul Wahab$~^\alpha$, Pascal Cotret$~^\beta$, Mounir Nasr Allah$~^\delta$\\Guillaume Hiet$~^\delta$, Arnab Kumar Biswas$~^\gamma$, Vianney Lapôtre$~^\gamma$, Guy Gogniat$~^\gamma$\\~\\
$~^\alpha$ IETR/SCEE research group, firstname.lastname@centralesupelec.fr\\
$~^\beta$ Independent researcher, pascal.cotret@gmail.com\\
$~^\delta$ INRIA/CIDRE research group, firstname.lastname@centralesupelec.fr\\
$~^\gamma$ Lab-STICC/University of South Brittany, firstname.lastname@univ-ubs.fr}
\title{\vspace{-4em}A novel lightweight hardware-assisted static instrumentation approach for ARM SoC using debug components}
\date{(Preprint version)}
\begin{document}
\maketitle

\begin{abstract}
	Most of hardware-assisted solutions for software security, program monitoring, and event-checking approaches require instrumentation of the target software, an operation which can be performed using an SBI (\emph{Static Binary Instrumentation}) or a DBI (\emph{Dynamic Binary Instrumentation}) framework. Hardware-assisted instrumentation can use one of these two solutions to instrument data to a memory-mapped register. Both these approaches require an in-depth knowledge of frameworks and an important amount of software modifications in order to instrument a whole application. This work proposes a novel way to instrument an application with minor modifications, at the source code level, taking advantage of underlying hardware debug components such as CS (\emph{CoreSight}) components available on Xilinx Zynq SoCs. As an example, the instrumentation approach proposed in this work is used to detect a double free security attack. Furthermore, it is evaluated in terms of runtime and area overhead. Results show that the proposed solution takes 30 $\mu$s on average to instrument an instruction while the optimized version only takes 0.014 $\mu$s which is ten times better than usual memory-mapped register solutions used in existing works \cite{Heo2015,Wahab2017}.
\end{abstract}

\section{Introduction}\label{introduction}
Software security is still a hot topic despite an important amount of research and development. Existing solutions are either too expensive in terms of cost, performance, power and area or either target a limited threat model. On the other side, attackers have more and more tools and vulnerabilities available in order to exploit existing systems. Therefore, it is important to provide solutions that can be easily adjusted on existing embedded systems without requiring important development effort.\\ 

One common approach for software security is performed through events monitoring (such as library calls, syscalls, specific instructions and so on). However, software-only solutions add important runtime overhead \cite{Dalton2007}: that is the reason why hardware-assisted solutions have been proposed. While they improve the performance overhead, proof-of-concepts are usually implemented on FPGAs rather than heterogeneous SoCs due to the amount of effort, time and money required to develop secure solutions on these platforms. Therefore, most of existing works cannot retrieve information required for monitoring purposes on a hardcore CPU.\\

Instrumentation is basically a transformation of a program into its own measurement tool and is commonly used in hardware-assisted software security solutions. It is used in DIFT (\emph{Dynamic Information Flow Tracking}) \cite{Wahab2017} in order to protect against different types of software attacks. Instrumentation can also be used for behavior monitoring \cite{Drewry_Flayer}, performance analysis \cite{Shende_tau} and software error detection \cite{error_detection_instrumentation}. As it was previously written, instrumentation can be done statically or dynamically. Static approaches modify the binary without requiring another process and cover all paths of the code while dynamic solutions require another process that instruments the binary and provides information only on the path taken by the application. Static approaches provide less information than dynamic ones but are usually faster in terms of performance because they do not require information at runtime.\\

All existing works using instrumentation do not provide a detailed description of their implementations. Some works such as \cite{Heo2015} use custom tools without describing them while others use compilers or existing dynamic instrumentation frameworks for instrumentation without providing in-depth details. RevARM \cite{Kim2017} is an instrumentation framework for ARM binaries which is oversized for this work.\\ 

This paper puts forward a novel approach for static instrumentation that can be used on ARM SoCs with CS (\emph{CoreSight}) debug components \cite{ARMCoreSightTRM}. This work does not target some features of instrumentation frameworks allowing to place the code at a given location. Nevertheless, it presents how the code is instrumented and how the instrumented data can be recovered on a reconfigurable device such as those included in Xilinx Zynq SoCs. The main goal of this work is to propose an instrumentation solution for ARM-based SoCs which is easy to implement with minor modifications, targeting modern OS (\emph{operating system}) such as Linux kernel. Furthermore, the approach developed in this work is able to send both user space and kernel space information.\\ 

This paper is organized as follows. Section \ref{related-work-and-assumptions} provides insights on existing instrumentation approaches. Section \ref{proposed-approach} presents the proposed architecture and provides implementation details. Section \ref{case-studies} provides different use cases. Section \ref{implementation-results} details implementation results and Section \ref{conclusion} gives some conclusions and future perspectives.

\section{Related work and assumptions}\label{related-work-and-assumptions}
A lot of works have been done on instrumentation frameworks: these frameworks can be architecture-dependent or independent, secure, static or dynamic. 
Soot \cite{soot} is a static instrumentation framework for Java and Android applications. Wala \cite{wala} also provides static instrumentation library for Java bytecode. Atom \cite{Srivastav_atom} is another static binary instrumentation framework on the Alpha processor platform for the Tru-64 OS. PEBIL \cite{PEBIL} is a static binary instrumentation framework for x86-64 architecture. Dyninst \cite{DynInst} is a static and dynamic instrumentation framework for multiple platforms (mainly x86-64 and ARM). Hijacker \cite{Pellegrini2013} is an open-source customizable static binary instrumentation tool. CSI (\emph{Comprehensive Static Instrumentation}) \cite{CSI_LLVM} is also a static instrumentation framework for LLVM. LLVM \cite{llvm}, an open-source compiler, can be used to create passes and instrument code.\\  

In \cite{Fogarty2012}, the author presented an original solution using the debug components and a secondary CPU core (based on the NXP CPU12X architecture) to extract instrumentation data. The main drawback of this solution is that it wastes another GPP (\emph{General Purpose Processor}) for instrumentation: as a consequence, the power consumption of this solution is doubled.\\  

Hardware-assisted instrumentation can use one of the above methods to instrument the application in order to recover instrumented data on the FPGA part of a heterogeneous SoC. However, using these frameworks require an in-depth knowledge of their API. In modern OS (e.g. Linux), the memory-mapped solution used in \cite{Heo2015,Wahab2017} requires an important number of software modifications. More precisely, it requires to map the physical address of the instrumentation register to a virtual address. This modification requires changing the kernel ELF binary loader. Then, this virtual address is sent towards the register included in instrumented instructions (which can be done through relocation). Finally, the binary can use this virtual address to write instrumented data. These changes require invasive modifications of the kernel. The solution proposed in this work provides the lowest amount of software modifications.\\ 

\noindent The contributions of this work are the following: 
\begin{itemize}
	\item A novel, simple and hardware-assisted instrumentation approach that takes advantage of CS debug components.
	\item This work proposes to add a system call to take advantage of the context ID feature provided by the CS PTM component and use the CS TPIU component and the EMIO interface in order to send trace and instrumented data towards the FPGA part.
	\item An improved and the first open-source version of the PFT (\emph{Program Flow Trace}) decoder that allows to decode trace and recover instrumented data on the FPGA part of ARM SoC such as Xilinx Zynq. 
\end{itemize}

\section{Proposed approach}\label{proposed-approach}
\subsection{Software modifications}
Instrumenting code using CS components can be done with a specific configuration; then, by adding a syscall and using the newly added syscall with a few lines of C code.\vspace{-.5em}

\subsubsection{Configuring Coresight}\label{subsubsec:cs_configuration}

ARM CS components technical reference manual \cite{ARMCoreSightTRM} explains how to program all CS components. The Linux kernel 4.9 provides a driver for CS components. However, the support for CS components on Zynq SoC was missing. The device tree was patched in order to use Linux kernel drivers of CS components.

The approach proposed in this work requires the activation of a specific feature known as context ID, that has not been used in any existing work to the best of our knowledge. Enabling context ID generates specific PFT (\emph{Program Flow Trace}) packets providing information about the context ID of an application. The context ID of an application consists of the PID (\emph{Process ID}) and ASID (\emph{Application Specific ID}) of the application.\\

The kernel is responsible for writing this value to a specific context id register. By enabling context ID tracing in the CS PTM component, this value is sent into the trace. Instead of writing PID and ASID into the context ID register, the value to be written is the value to instrument.\newpage

\begin{table}[htbp]
	\begin{center}
		\caption{CS configuration in Linux kernel driver}
		\begin{tabular}{|l|c|l|}
			\hline
			\textbf{File name} & \textbf{Value} & \textbf{Description}\\ \hline%
			\texttt{mode} & 0x30 &\begin{tabular}[l]{@{}l@{}}Enable branch broadcast\\and context ID feature\end{tabular}\\ \hline
			\texttt{addr\_idx} & 1 & Choose address comparator \\ \hline
			\texttt{addr\_acctype} & 0 & Choose comparison access type\\ \hline
			\texttt{addr\_range} & \begin{tabular}[c]{@{}c@{}}\textbf{0x106a0}\\\textbf{0x10700}\end{tabular} & \begin{tabular}[l]{@{}l@{}}Enable trace in\\address space given\end{tabular}\\ \hline
			\texttt{enable\_source} & 1 & Enable PTM component \\ \hline
		\end{tabular}
		\label{tab:CSConfiguration}
	\end{center}
\end{table}

Table \ref{tab:CSConfiguration} shows the values used to program the PTM component in the Linux kernel sysfs file system. TPIU (\emph{Trace Port Interface Unit}) is used as trace sink to recover trace on the FPGA part via EMIO interface \cite{Wahab2017}.  

\subsubsection{Adding syscall}

Listing \ref{lst:custom_syscall_code} shows the kernel code to be added in order to write the value to the context ID register. The \texttt{mcr} instruction is used to write the instrumented value to the context ID register located in coprocessor 15 while the \texttt{isb()} instruction triggers the write \cite{ARMArchiManual}. 
\begin{lstlisting}[language=C, caption={Custom syscall code in \texttt{kernel/sys.c} file},label={lst:custom_syscall_code}]
SYSCALL_DEFINE1(wctxtid, unsigned int, instrumentation_data)
{
  asm volatile("mcr p15, 0, %0, c13, c0, 1\n"::"r"(instrumentation_data):);  
  isb();
  return 0;
}
\end{lstlisting}

The syscall definition needs to be added into the kernel source as shown in Listing \ref{lst:custom_syscall_definition}.
\begin{lstlisting}[language=C, caption={Custom syscall definition in \texttt{include/linux/syscalls.h} file},label={lst:custom_syscall_definition}]
asmlinkage long sys_wctxtid (unsigned int instrumentation_data);
\end{lstlisting}

Finally, a number needs to be associated with the new syscall as shown in Listing \ref{lst:number_custom_syscall}.
\begin{lstlisting}[language=C, caption={Associate a number to custom syscall in \texttt{arch/arm/kernel/calls.S} file},label={lst:number_custom_syscall}]
/* 397 */   CALL(sys_wctxtid)
\end{lstlisting}

\subsubsection{Writing C code}

Once the kernel code has been modified, it is compiled with the write to the context ID register option enabled in order to obtain the Linux kernel binary file (\texttt{uImage}). Once the modified kernel is booted, Listing \ref{lst:instrumented_code} shows how the newly added syscall can be used in any C program.\newpage 
\begin{lstlisting}[language=C, caption={Example instrumented code},label={lst:instrumented_code}]
#include <unistd.h>
#include <sys/syscall.h>
int main()
{
  unsigned dataToSend = 0x1234abcd;
  // system call number is 397
  // (cf calls.S file change)
  syscall(397, dataToSend);
  return 0;
}
\end{lstlisting}

\subsection{Hardware design}

\subsubsection{Overall architecture}
\begin{figure}[htbp]
	\centering
	\includegraphics[width=.72\linewidth, keepaspectratio]{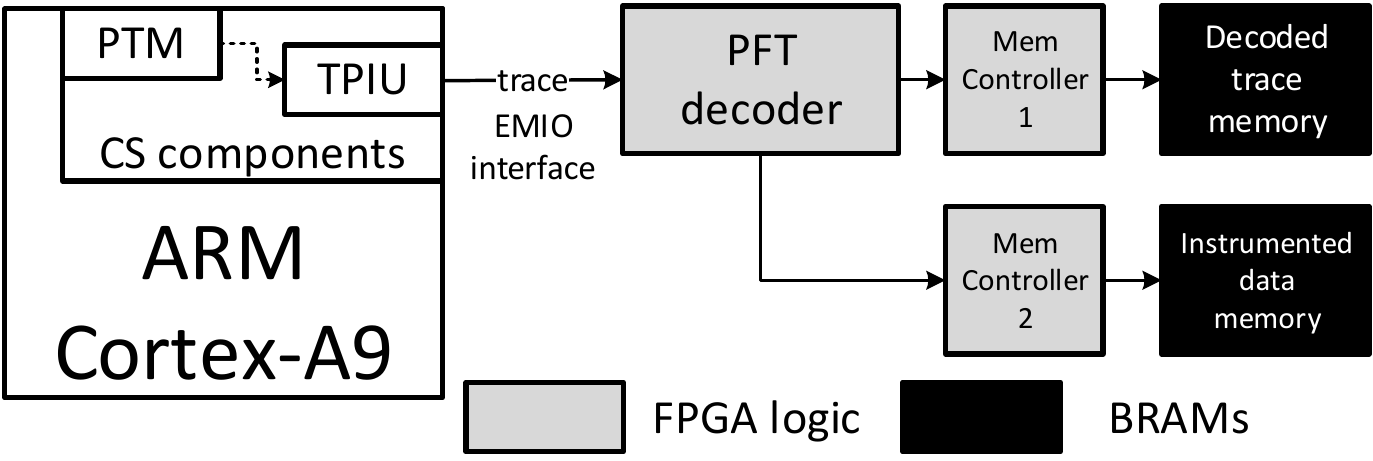}
	\caption{Overall architecture}
	\label{fig:overall_architecture}
\end{figure}

Figure \ref{fig:overall_architecture} shows the overall architecture of the approach proposed in this work (implemented on a Zynq SoC) with a threat model similar to \cite{Wahab2017} where main threats come from the CPU and the link between the CPU and the FPGA logic.

After configuring CS components and running the program, trace is obtained on the FPGA part through the EMIO interface according to the PFT (\emph{Program Flow Trace}) protocol \cite{ARMCoreSightPFT}. The important module developed in this work is the PFT decoder that differs with previous existing implementation. Furthermore, this work improves the solution of Wahab et al. \cite{Wahab2017} by adding the support of the context ID register. The entire design and code for this work is available on Bitbucket: \texttt{https://bitbucket.org/hardblare/}\\\texttt{hw-static-instrumentation}.

\subsubsection{PFT decoder}

\begin{figure}[htbp]
	\centering
	\includegraphics[width=\linewidth, keepaspectratio]{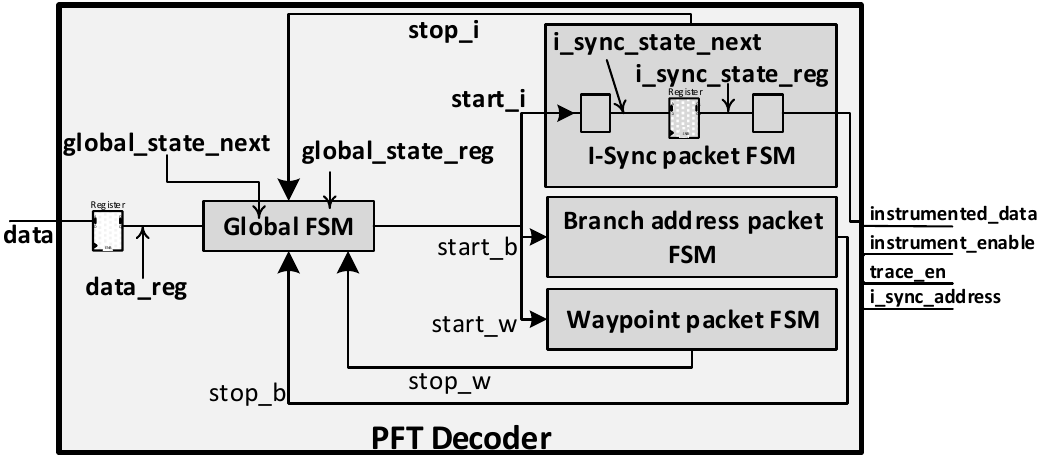}
	\caption{PFT decoder architecture}
	\label{fig:pft_decoder}
\end{figure}

The TPIU export raw traces to the FPGA part via EMIO interface. Listing \ref{lst:raw_trace} shows an example of raw trace. Trace must be decoded in order to recover instrumented data on the FPGA part. The PFT protocol \cite{ARMCoreSightPFT} presents 11 different trace packets and their corresponding headers. Figure \ref{fig:pft_decoder} shows the overall architecture of the PFT decoder. The PFT decoder consists of four FSMs (\emph{Finite State Machines}): a global FSM that controls the other three packet FSMs (I-Sync, branch address, and waypoint). There are other PFT packets such as a-sync, exception that are also decoded by the global FSM. The most important PFT packet in this work is the I-Sync packet as it contains the instrumented data. The global state machine detects packet type and enables the corresponding packet FSM by setting start signal.

\begin{figure}[htbp]
	\centering
	\includegraphics[width=.53\linewidth]{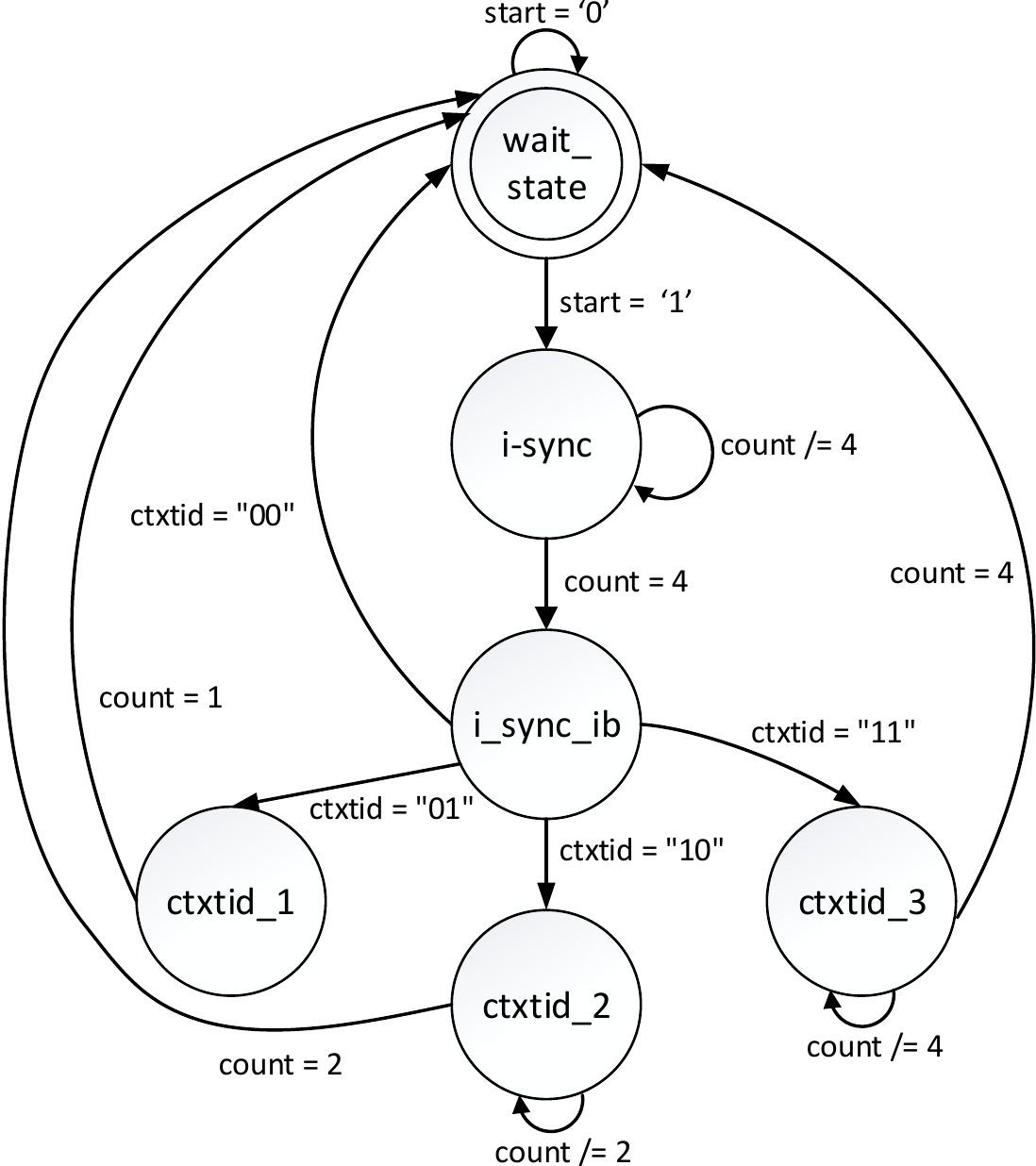}
	\caption{Finite State Machine diagram for the I-Sync packet decoder}
	\label{fig:state_machine_i-sync}
\end{figure}

Figure \ref{fig:state_machine_i-sync} shows the FSM diagram of I-Sync packet. The FSM is in \texttt{wait\_state} by default. When the start signal is set by the global FSM, I-Sync packet FSM goes into \texttt{i\_sync} state and starts counting trace samples. The received four samples contain the current PC (\emph{Program Counter}) value of ARM CPU. Once four samples are received, the state machine goes into \texttt{i\_sync\_ib} state. If context id tracing is enabled, it can send one, two or four bytes. Depending on the generic \texttt{ctxtid} value, the context id packet is decoded. For instance, if the generic \texttt{ctxtid} is equal to ``11'', then the FSM starts a counter that counts the number of received trace samples.

\newpage

When this number is equal to 4, then the instrumented data bytes are correctly received. When the packet FSM finishes decoding packet, it goes back to \texttt{wait\_state} and sends the stop signal to global FSM which then looks for the next packet type. The other packet FSMs use similar mechanism to communicate with global FSM and have similar state machines.

\begin{figure}[htbp]
	\centering
	\includegraphics[width=\linewidth]{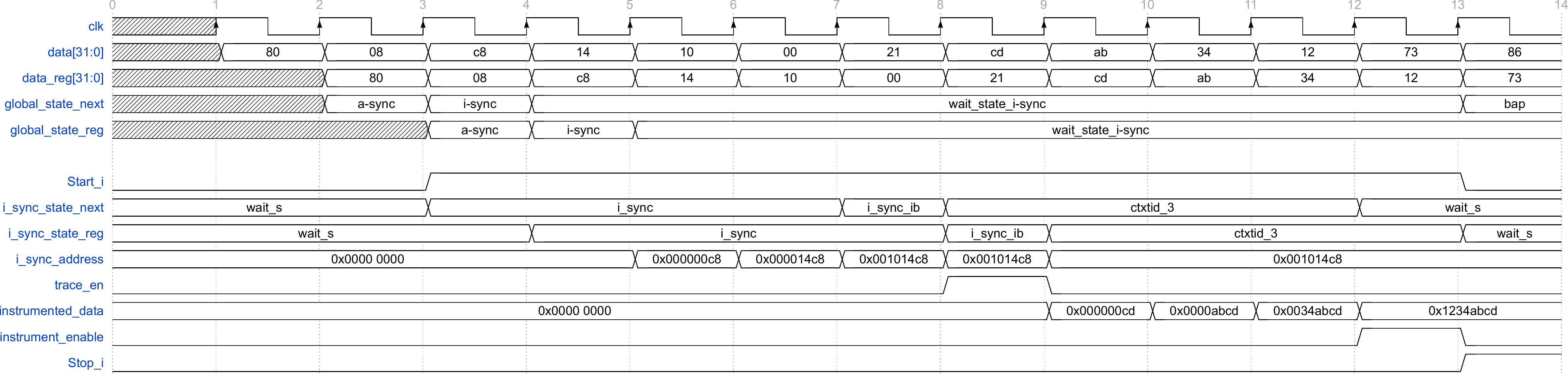}
	\caption{Timing diagram of global FSM and I-Sync FSM of PFT decoder (Figure \ref{fig:pft_decoder})}
	\label{fig:pft_decoder_fsms_timing}
\end{figure}

Figure \ref{fig:pft_decoder_fsms_timing} is a timing diagram showing how the PFT decoder works at each clock cycle. All inputs are registered: for instance, input \texttt{data} is registered to obtain \texttt{data\_reg} signal. All FSMs work with \texttt{data\_reg} registered signal. When the global FSM detects an I-Sync packet, it enables a \texttt{start\_i} signal (e.g. this is the case at the third clock cycle) to enable the corresponding slave FSM. The I-Sync packet FSM then decodes the packet according to the PFT architecture as shown by \texttt{i\_sync\_state\_reg}, \texttt{i\_sync\_state\_next}, \texttt{i\_sync\_address}, \texttt{instrumented\_data} and \texttt{instrument\_en} signals. The \texttt{instrumented\_data} signal contains the value instrumented by software. In the meantime, the global FSM waits for a stop signal which is enabled by the packet FSM when the packet is decoded. Enabling \texttt{stop\_i} signal at the thirteenth clock cycle changes the global FSM state according to \texttt{data\_reg} signal. This allows to decode the received trace on-the-fly. The \texttt{trace\_en} signal allows to determine when valid decoded trace is available while the \texttt{instrument\_enable} signal allows to determine when valid instrumented data is available. Outputs of the PFT decoder are also registered to avoid timing failures due to longer critical paths.\\

Listing \ref{lst:raw_trace} shows the trace at the input of PFT decoder. It contains different packets detailed in a Technical Reference Manual provided by ARM \cite{ARMCoreSightPFT}.
\begin{lstlisting}[language=bash,style=mystyle,caption={Raw trace},label={lst:raw_trace}]
root@zedboard-hardblare:~/tests_appli#./trace_etb
00000000  <@\textcolor{red}{00 00 00 00 00 80}@> <@\textcolor{ForestGreen}{ 08 78}@>
<@\textcolor{ForestGreen}{04 01 00 21 \textbf{f4 ee 03 00}}@>
00000010  <@\textcolor{blue}{8b 03}@> <@\textcolor{ForestGreen}{08 8c 04 01 00 21}@>
<@\textcolor{ForestGreen}{\textbf{f4 ee 03 00}}@> <@\textcolor{blue}{9d 03}@> <@\textcolor{ForestGreen}{08 98}@>
00000020  <@\textcolor{ForestGreen}{04 01 00 21}@> <@\textcolor{ForestGreen}{\textbf{ff ff ff ff}}@>
<@\textcolor{blue}{9d 03}@> <@\textcolor{ForestGreen}{08 a8 04 01 00 21}@>
00000030  <@\textcolor{ForestGreen}{\textbf{dd dd dd dd}}@> <@\textcolor{blue}{85 03}@> <@\textcolor{ForestGreen}{08 b4}@>
<@\textcolor{ForestGreen}{04 01 00 21}@> <@\textcolor{ForestGreen}{\textbf{dd dd dd dd}}@>
00000040  <@\textcolor{blue}{9d 03}@> <@\textcolor{ForestGreen}{08 c4 04 01 00 21}@>
<@\textcolor{ForestGreen}{\textbf{aa aa aa aa}}@> <@\textcolor{blue}{9d 03}@> <@\textcolor{ForestGreen}{08 d4}@>
00000050  <@\textcolor{ForestGreen}{04 01 00 21}@> <@\textcolor{ForestGreen}{\textbf{11 11 11 11}}@>
<@\textcolor{blue}{fd bc cf db 0d 01}@> <@\textcolor{red}{00 00}@>
<@\textcolor{ForestGreen}{\textbf{PFT packets}}@>
<@\textcolor{red}{a-sync packet}@> <@\textcolor{ForestGreen}{i-sync packet}@> <@\textcolor{blue}{branch address packet}@>
\end{lstlisting}

Listing \ref{lst:decoded_trace} shows the content of the instrumented data memory. The obtained values correspond to the values inserted in the program source code.
\begin{lstlisting}[language=bash,style=mystyle,caption={Instrumented data from decoded raw trace},label={lst:decoded_trace}]
<@\bfseries0003eef4@> <@\bfseries0003eef4@> ffffffff
dddddddd aaaaaaaa 11111111
\end{lstlisting}

\subsection{Optimization}\label{optimization}
Rather than creating a new syscall as proposed earlier, this work proposes to modify an existing syscall called in the instrumented application. For example, if a \texttt{malloc} call is done in the instrumented program and its behavior needs to be monitored, the associated syscall , which is either \texttt{mmap} or \texttt{brk}, can be modified in order to add the proposed code inside the existing syscall. This way, a specific syscall is not required while reducing the runtime overhead.\\

This optimization is not possible if the program does not use any syscall which is rather rare as most programs require kernel services which are only accessible through syscalls. Another optimization could be to send multiple instrumented data values at each syscall rather than sending one value. However, this is not possible due to the fact that the PTM sends only the last instrumented value rather than all instrumented values during a syscall. 

\section{Case studies}\label{case-studies}
\subsection{Double Free}\label{double-free}
Listing \ref{lst:double_free_example_code} shows a straightforward double-free attack. There are three allocated memory areas: A, B and C. First, A and C are allocated. Then, A is freed. Later, B is allocated and A is freed again. This is the line where the attack happens.
However, when this code is compiled and executed, the Linux kernel does not detect any error. This type of attack can lead to heap overflow, attack code execution or illegitimate privilege elevation. 
\begin{lstlisting}[language=C, caption={Double free vulnerability example code},label={lst:double_free_example_code}]
#include <stdio.h>
#include <stdlib.h>
#include <string.h>

#define WORD_SIZE (size_t)-1>0xFFFFFFFFUL?8:4
#define SIZEOFNORMALCHUNK 0x100-WORD_SIZE
#define SIZEOFFASTCHUNK 0x60-WORD_SIZE

int main()
{
  char *A, *B, *C;
  A = malloc(SIZEOFNORMALCHUNK);
  C = malloc(SIZEOFNORMALCHUNK);
  free(A);
  // same location as A
  B = malloc(SIZEOFNORMALCHUNK); 
  if (B)
    // Double Free!
    free(A);
  return 0;
}
\end{lstlisting}

Listing \ref{lst:double_free_instrumented_code} shows the instrumented code in order to detect the double free attack. Each time a \texttt{malloc} or \texttt{free} function is called, another instruction is added in order to send the information about the library function. In order to distinguish a \texttt{malloc} call from a \texttt{free} call, an example value of 12 MSB bits (\texttt{0xfff}) is added to the \texttt{malloc} call whereas \texttt{(0xffe)} is added to the \texttt{free} library call. Then, the lower 20 bits are used to specify the region allocated or freed. As an example, numbers are shown in instrumented value but static analysis can provide generic variables, containing numerical values, that can be easily inserted instead of numbers.
\begin{lstlisting}[language=C, caption={Instrumented code for double free detection},label={lst:double_free_instrumented_code}]
A = malloc(SIZEOFNORMALCHUNK);
// instrumented code
syscall(397,0xfff00001); 
C = malloc(SIZEOFNORMALCHUNK);
// instrumented code
syscall(397,0xfff00002); 
free(A);
// instrumented code
syscall(397,0xffe00001);
// same location as A  
B = malloc(SIZEOFNORMALCHUNK); 
// instrumented code
syscall(397,0xfff00003); 
if (B){
  free(A); 
  // instrumented code
  syscall(397,0xffe00001); 
}
\end{lstlisting}

Then, the program is compiled and executed after configuring CS components as explained in subsection \ref{subsubsec:cs_configuration}. An example output of the program execution and the content of memories is shown in Listing \ref{lst:double_free_execution_output}.\\

After the configuration of CS components, the program is launched and the content of the decoded trace memory is recovered using AXI BRAM controller (not shown in Figure \ref{fig:overall_architecture} as it is not required for the proposed design but is added for debug purposes). The first value obtained in the decoded trace is \texttt{0x106a0} which is the address where tracing is started as configured in CS components (see section \ref{subsubsec:cs_configuration}).\\

Then, next values show the control flow changes in the program. In this case study, the decoded trace is not used. Then, the content of the instrumented data memory is displayed (containing the values inserted into the program). It can be noticed that the value \texttt{0xffe00001} is present twice while \texttt{0xfff00001} is present only once which means that the first region is freed twice and allocated once. This example shows that the instrumented data can be successfully recovered on the FPGA part and that the double free attack can be detected.\newpage 
\begin{lstlisting}[language=bash,caption={Execution output for double free example code}, label={lst:double_free_execution_output}]
<@\textbf{root@zedboard:~/tests-appli\#
./trace-tpiu-topleaks}@>

coresight-tpiu f8803000.tpiu:
TPIU enabled
coresight-replicator amba:replicator:
REPLICATOR enabled
coresight-funnel f8804000.funnel:
FUNNEL inport 0 enabled
coresight-etm3x f889c000.ptm0:
ETM tracing enabled

DECODE TRACE
00 106a0 10358 106c0 104d4 106ec b6e3ec88 1057c 
10378 10598 1039c 105a0 103c0 105b8 1039c 105c0
103c0 105d8 10384 105e0 103c0 105f0 10378 10600 
1039c 10608 103c0 10620 10384 10628 103c0 10638
10390 10644 10378 10654 10378 10660 1050c 1066c 
10390 10678 10378 10690 b6e3ecf8 b6e3ec00 00 00

<@\textbf{root@zedboard:~/tests-appli\#}@>
<@\textbf{./get-instrumented-data.elf}@>
/dev/mem opened.
Memory mapped at address 0xb6fba000.
fff00001 fff00002 ffe00001 fff00003 
ffe00001 00 00 00 00 00 00 00 00 00
\end{lstlisting}

\subsection{Other use cases}
This method allows reconstructing the CFG on the FPGA side using a decoded trace \cite{Wahab2017}. Therefore, using  decoded trace, a control flow checker unit can be implemented on the FPGA to make sure that no control-flow variations occur during program execution. This information is very important as it cannot be determined statically. Moreover, it can be recovered without adding syscall by correctly configuring CS components and avoiding important runtime overhead in existing control-flow checker solutions.\\

There are other attacks that can be detected using this instrumentation strategy (for instance, event checking): static analysis can detect events that are sent towards the FPGA part in order to verify some properties. Therefore, this work suits the LTL (\emph{Linear Temporal Logic}) hardware-assisted verification solutions. 

\section{Implementation Results}\label{implementation-results}
Xilinx tools 2017.1 are used on a Xilinx Zedboard with a Z-7020 SoC (dual-core Cortex-A9 running at 667 MHz and an Artix-7 FPGA) to implement the architecture shown in Figure \ref{fig:overall_architecture}. The PFT decoder and memory controllers are working at the frequency of 250 MHz. Following subsections present an evaluation of this work in terms of runtime and area overheads.

\subsection{Time overhead analysis}\label{performance-overhead}
CS components do not add any execution time overhead as shown by Wahab et al. \cite{Wahab2017}. Even though this work uses a different configuration of CS components (context ID tracing is enabled), CS components still do not affect the execution time. The time overhead is measured by using Linux \texttt{perf} command and by launching MiBench benchmark applications with and without enabling CS components.\\ 

The overhead is only due to syscalls. On the target platform, the overhead of added syscall is 30 $\mu$s while being 0.150 $\mu$s for the memory-mapped register approach. This work presents a method which is slower than existing memory-mapped approaches. However, with the optimization (subsection \ref{optimization}), time overhead can be reduced. If additional instructions \texttt{mcr} and \texttt{isb} are added to an existing call, the overhead due to the syscall is removed and the global overhead is only due to the execution of \texttt{mcr} and \texttt{isb} instructions. The \texttt{mcr} instruction takes $1+x$ cycles (where $x$ is the number of cycles spent in the coprocessor busy-wait loop) and the \texttt{isb} instruction takes 4 cycles. Therefore, both instructions take $1+x+4 = x+5$ cycles at 667MHz. If this value is equal to ten cycles which is the case if the coprocessor is availble, then the runtime overhead introduced would be 0.014 $\mu$s which is more than ten times better than the 0.15 $\mu$s required for the memory-mapped register approach.

\subsection{Latency and bandwidth}
If \texttt{n} is the number of packets received in a PFT packet, the PFT decoder requires (\texttt{n}+1) clock cycles to decode a trace. The maximum value of \texttt{n} in this work is 10. In other words, the PFT decoder requires only one additional cycle to decode a trace and recover instrumented data while the solution based on memory-mapped register requires on average 30 cycles in average to receive valid instrumented data due to timing delays introduced by the handshake between different channels. In this work, as 8 bits are sent at each clock cycle at the frequency of 250 MHz, the trace bandwidth is 250*8 = 2000 Mbits/s. The maximum bandwidth of the trace port is 250*32 = 8000 Mbits/s. The AXI interface on a Zedboard can go up to 200 MHz. Therefore, the maximum bandwidth of the AXI port is 200*32 = 6400 Mbits/s. 

\subsection{Area results}\label{area-results}
Table \ref{tab:results} shows the area overhead of proposed architecture which only requires 0.34\% of slice LUTs, 0.47\% of slice registers and 2.86\% of BRAMs on a Zynq Z-7020 device. The memory-mapped solution requires 1684 slice LUTs (3.17\%), 1523 slice registers (1.43\%) and no BRAM. The memory-mapped register solution takes more area because the memory required to store data is implemented in FPGA itself rather than in BRAM as in this work. Furthermore, the memory-mapped register approach requires an AXI interconnect block which has internal FIFOs making it larger in terms of area occupation.\newpage

\begin{table}[htbp]
	\begin{center}
		\caption{Post-implementation area results of overall architecture and memory-mapped approach on Xilinx Zynq Z-7020}
			\begin{tabular}{|c|c|c|c|}
				\hline
				\textbf{IP Name} & \begin{tabular}[c]{@{}c@{}}\textbf{Slice LUTs}\\ \textbf{(in \%)}\end{tabular} & \begin{tabular}[c]{@{}c@{}}\textbf{Slice registers}\\ \textbf{(in \%)}\end{tabular}  & \textbf{BRAM tiles} \\ \hline
				PFT Decoder & 126 (0.24\%) & 240 (0.23\%) & 0 \\  \hline 
				Mem controller 1 & 18 (0.03\%) & 79 (0.08\%) & 0 \\  \hline
				Mem controller 2 & 18 (0.03\%) & 79 (0.08\%) & 0 \\  \hline
				\begin{tabular}[c]{@{}c@{}}Decoded\\ trace memory\end{tabular} & \begin{tabular}[c]{@{}c@{}}2\\ (0.01\%)\end{tabular} & 0 & \begin{tabular}[c]{@{}c@{}}2\\ (1.43\%)\end{tabular} \\ \hline 
				\begin{tabular}[c]{@{}c@{}}Instrumented\\ data memory\end{tabular} & \begin{tabular}[c]{@{}c@{}}2\\ (0.01\%)\end{tabular} & 0 & \begin{tabular}[c]{@{}c@{}}2\\ (1.43\%)\end{tabular} \\ \hline 
				Miscellaneous & 16 (0.03\%) & 97 (0.09\%) & 0 \\ \hlineB{2}
				\multirow{1}{*}{\textbf{Total Design}} & \textbf{182 (0.34\%)} & \textbf{495 (0.47\%)} & \textbf{4 (2.86\%)} \\ \hline \hline
				\multirow{1}{*}{\textbf{Memory-mapped}} & \textbf{1684 (3.17\%)} & \textbf{1523 (1.43\%)} & \textbf{0 (0\%)} \\ \hline
				\textbf{Total Available} & \textbf{53200} & \textbf{106400} & \textbf{140} \\ \hline
			\end{tabular}
		\label{tab:results}
	\end{center}
\end{table}

\subsection{Comparison with related works}\label{comparison-previous-works}
Table \ref{tab:comparison} compares this work with the usual memory-mapped register solution which is widely used in existing hardware-assisted instrumentation approaches. This work requires only few minor modifications of the kernel (addition of \texttt{mcr} and \texttt{isb} instructions) while memory-mapped register solutions need a modified kernel ELF loader and relocation units. This work has a lower latency and a higher maximum bandwidth than the memory-mapped solution. Furthermore, the area overhead is 6 times smaller than existing works. 

\begin{table}[htbp]
	\begin{center}
		\caption{Comparison with previous works}
			\begin{tabular}{|c|c|c|c|}
				\hline
				\textbf{Metric} & \begin{tabular}[c]{@{}c@{}}\textbf{Solution}\\ \textbf{Section III.A}\end{tabular} & \begin{tabular}[c]{@{}c@{}}\textbf{Solution}\\ \textbf{Section III.C}\end{tabular} & \begin{tabular}[c]{@{}c@{}}\textbf{Existing works}\\ \textbf{(memory-mapped)}\end{tabular} \\ \hline%
				\begin{tabular}[c]{@{}c@{}}Software\\ modifications\end{tabular} & \textbf{low} & \textbf{low} & moderate \\ \hline
				\begin{tabular}[c]{@{}c@{}}Latency\\ (clock cycles)\end{tabular} & (\texttt{n}+1) $\leq$ 10 & (\texttt{n}+1) $\leq$ 10 & 30 \\ \hline
				\begin{tabular}[c]{@{}c@{}}Maximum bandwidth\\ (Mbits/s)\end{tabular} & \textbf{8000} & \textbf{8000} & 6400 \\ \hline
				\begin{tabular}[c]{@{}c@{}}Runtime\\overhead ($\mu$s)\end{tabular} & 30 & \textbf{0.014} & 0.150 \\ \hline
				\begin{tabular}[c]{@{}c@{}}Area\\overhead (\%)\end{tabular} & \textbf{0.47} & \textbf{0.47} & 3.17 \\
				\hline
			\end{tabular}
		\label{tab:comparison}
	\end{center}
\end{table}

\section{Conclusion}\label{conclusion}
This work proposes to exploit the context ID feature of the CoreSight PTM component in order to instrument an application. The generated trace, which also contains the instrumented data, is sent to the FPGA part using the CS TPIU component and EMIO interface. Software modifications required to instrument are minimal (about 10 lines are changed in the Linux kernel). Furthermore, the hardware design required to decode traces takes less than 0.5\% of the FPGA area on a Zynq SoC. This work takes 30 $\mu$s for each instruction added in the source code. The optimized version, which consists in removing context switch time overhead by modifying existing syscall used in the application rather than adding a new syscall, give a runtime overhead of 0.014 $\mu$s: it is 10 times better than 0.15 $\mu$s required for a memory-mapped register solution. In terms of perspectives, the approach presented in this work will be ported to Intel-based SoCs by taking advantage of the Intel PT (\emph{Processor Trace}) debug component.

\bibliographystyle{unsrt}
\bibliography{IEEEabrv}

\begin{thebibliography}{10}

\bibitem{Dalton2007}
Michael Dalton, Hari Kannan, and Christos Kozyrakis.
\newblock Raksha: A flexible information flow architecture for software
  security.
\newblock In {\em Proceedings of the 34th Annual International Symposium on
  Computer Architecture}, ISCA '07, pages 482--493, New York, NY, USA, 2007.
  ACM.

\bibitem{Wahab2017}
M.~A. Wahab, P.~Cotret, M.~N. Allah, G.~Hiet, V.~Lapôtre, and G.~Gogniat.
\newblock Armhex: A hardware extension for dift on arm-based socs.
\newblock In {\em 2017 27th International Conference on Field Programmable
  Logic and Applications (FPL)}, pages 1--7, September 2017.

\bibitem{Drewry_Flayer}
Will Drewry and Tavis Ormandy.
\newblock Flayer: Exposing application internals.
\newblock In {\em Proceedings of the First USENIX Workshop on Offensive
  Technologies}, WOOT '07. USENIX Association, 2007.

\bibitem{Shende_tau}
Sameer~S. Shende and Allen~D. Malony.
\newblock The tau parallel performance system.
\newblock {\em The International Journal of High Performance Computing
  Applications}, 20(2):287--311, 2006.

\bibitem{error_detection_instrumentation}
Sunsup So, Yongseop Lim, Sung~Deok Cha, and Yong~Rae Kwon.
\newblock An empirical study on software error detection: voting,
  instrumentation and fagan inspection.
\newblock In {\em Proceedings 1995 Asia Pacific Software Engineering
  Conference}, pages 345--351, December 1995.

\bibitem{Heo2015}
Ingoo Heo, Minsu Kim, Yongje Lee, Changho Choi, Jinyong Lee, Brent~Byunghoon
  Kang, and Yunheung Paek.
\newblock Implementing an application-specific instruction-set processor for
  system-level dynamic program analysis engines.
\newblock {\em ACM Trans. Des. Autom. Electron. Syst.}, 20(4):53:1--53:32,
  September 2015.

\bibitem{Kim2017}
Taegyu Kim, Chung~Hwan Kim, Hongjun Choi, Yonghwi Kwon, Brendan Saltaformaggio,
  Xiangyu Zhang, and Dongyan Xu.
\newblock {RevARM}: A platform-agnostic {ARM} binary rewriter for security
  applications.
\newblock In {\em Proceedings of the 33rd Annual Computer Security Applications
  Conference}, ACSAC 2017, pages 412--424, New York, NY, USA, 2017. ACM.

\bibitem{ARMCoreSightTRM}
{ARM}.
\newblock {\em {CoreSight Components Technical Reference Manual}}.

\bibitem{soot}
{Sable Research Group of McGill University}.
\newblock Soot - a framework for analyzing and transforming java and android
  applications.

\bibitem{wala}
{IBM T.J. Watson Research Center}.
\newblock T.j. watson libraries for analysis (wala).

\bibitem{Srivastav_atom}
Amitabh Srivastava and Alan Eustace.
\newblock Atom: A system for building customized program analysis tools.
\newblock {\em SIGPLAN Not.}, 39(4):528--539, April 2004.

\bibitem{PEBIL}
M.~A. Laurenzano, M.~M. Tikir, L.~Carrington, and A.~Snavely.
\newblock Pebil: Efficient static binary instrumentation for linux.
\newblock In {\em 2010 IEEE International Symposium on Performance Analysis of
  Systems \& Software (ISPASS 2010)(ISPASS)}, volume~00, pages 175--183, March
  2010.

\bibitem{DynInst}
Andrew~R. Bernat and Barton~P. Miller.
\newblock Anywhere, any-time binary instrumentation.
\newblock In {\em Proceedings of the 10th ACM SIGPLAN-SIGSOFT Workshop on
  Program Analysis for Software Tools}, PASTE '11, pages 9--16, New York, NY,
  USA, 2011. ACM.

\bibitem{Pellegrini2013}
A.~Pellegrini.
\newblock Hijacker: Efficient static software instrumentation with applications
  in high performance computing: Poster paper.
\newblock In {\em 2013 International Conference on High Performance Computing
  Simulation (HPCS)}, pages 650--655, July 2013.

\bibitem{CSI_LLVM}
Tao~B. Schardl, Tyler Denniston, Damon Doucet, Bradley~C. Kuszmaul,
  I-Ting~Angelina Lee, and Charles~E. Leiserson.
\newblock The csi framework for compiler-inserted program instrumentation.
\newblock {\em Proc. ACM Meas. Anal. Comput. Syst.}, 1(2):43:1--43:25, December
  2017.

\bibitem{llvm}
{University of Illinois}.
\newblock The llvm compiler infrastructure.

\bibitem{Fogarty2012}
P.~Fogarty.
\newblock Minimising the impact of software instrumentation using on-chip debug
  and a secondary cpu core.
\newblock In {\em Proceedings of the 2012 System, Software, SoC and Silicon
  Debug Conference}, pages 1--5, September 2012.

\bibitem{ARMArchiManual}
{ARM}.
\newblock {\em ARM Architecture Reference Manual - ARMv7-A and ARMv7-R
  edition}.

\bibitem{ARMCoreSightPFT}
{ARM}.
\newblock {\em {CoreSight Program Flow Trace Architecture Specification}}.

\end{thebibliography}
\end{document}